# Templated direct growth of ultra-thin double-walled carbon nanotubes

Lei Shi,*,[1] Jinquan Wei,[2] Kazuhiro Yanagi,[3] Takeshi Saito,[4] Kecheng Cao,[5] Ute Kaiser,[5] Paola Ayala,[1] and Thomas Pichler[1]

[1]*University of Vienna, Faculty of Physics, 1090 Wien, Austria*

[2]*Tsinghua University, State Key Lab of New Ceramics and Fine Processing, School of ftaterials Science and Engineering, Beijing 100084, P. R. China*

[3]*Tokyo ftetropolitan University, Department of Physics, 1-1 ftinami-Osawa, Hachiouji, Tokyo 192-0397, Japan*

[4]*National Institute of Advanced Industrial Science and Technology (AIST), Nanomaterials Research Institute, Tsukuba 305-8565, Japan*

[5]*Ulm University, Central Facility for Electron fticroscopy, Electron fticroscopy Group of ftaterials Science, Ulm 89081, Germany*

E-mail: lei.shi@univie.ac.at

### Abstract

Double-walled carbon nanotubes (DWCNTs) combined the advantages of multi-walled (MW-) and single-walled (SW-) CNTs can be obtained by transforming the precursors (*e.g.* fullerene, ferrocene) into thin inner CNTs inside SWCNTs as templates. However, this method is limited since the DWCNT yield is strongly influenced by the filling efficiency (depending on the type of the filled molecules), opening and cutting the SWCNTs, and the diameter of the host SWCNTs. Therefore, it cannot be applied to all types of SWCNT templates. Here we show a universal route to synthesize ultra-thin

DWCNTs via making SWCNTs stable at high temperature in vacuum. This method applies to different types of SWCNTs including metallicity-sorted ones without using any precursors since the carbon sources were from the reconstructed SWCNTs and the residue carbons. The resulting DWCNTs are with high quality and the yield of inner tubes is comparable to/higher than that of the DWCNTs made from the transformation of ferrocene/fullerene peapods.

Double-walled carbon nanotubes (DWCNTs), as two-layered multi-walled CNTs (MWCNTs), combine the advantages of both MWCNTs and single-walled CNTs (SWCNTs), e.g. great conductivity and flexibility as the SWCNTs; chemical/oxidative stability as the MWCNTs [1–3]. Also, DWCNTs present superior electrical and thermal transport, resulting in promising nanodevices. [4,5] Hence, increasing attention has been paid to their synthesis[6–11], separation[12–15], properties[16–18], and applications[2,19,20]. Previously, direct synthesis of DWCNTs has been achieved by chemical vapour deposition (CVD)[6,7,9,11] and the arc-discharge method[8,10,21]. However, these methods normally produce large DWCNTs with a wide diameter distribution. In order to get thin DWCNTs with narrow diameter distribution, so called peapods, e.g. fullerene/ferrocene filled SWCNTs, can be transformed into DWCNTs by annealing in vacuum normally below 1300 °C[22–24]. The disadvantage on the peapod method is that filling the precursor molecules like fullerene/ferrocene as the first step is complicated, time consuming, and in small amount. Especially, it cannot apply to the SWCNTs with small diameter.

Here we introduce a one-step method to produce small-diameter DWCNTs with narrow diameter distribution in large scale, i.e. high-temperature annealing of SWCNTs at the temperature between 1400 and 1500 °C. High-temperature annealing of SWCNTs in inert gas or vacuum is commonly applied to purify CNTs, *e.g.* removing residual metals and reducing the wall defects[25-27]. However, SWCNTs, especially thin and/or defective ones, are structurally reformed at high temperature. Depending on the temperature, SWCNTs can be enlarged during the annealing[28-30] and combined into thick multi-walled CNTs (MWCNTs) at higher temperatures[29-32]. Similarly, even the DWCNTs are much more stable than the SWCNTs, they also can be transformed during the annealing into carbon bicables or heterojunctions or triple-walled CNTs, and finally into huge MWCNTs due to a coalescence reaction between the adjacent DWCNTs [31,33-35]. We recently reported that linear carbon chains can be synthesized inside thin DWCNTs[11,36] by high-temperature annealing in vacuum. Inspired on this, we have investigated in depth on the high-temperature annealing of SWCNTs and we have observed that it is possible to form inner tubes inside SWCNTs (but not coalesced into large MWCNTs), resulting in a universal route to synthesize DWCNTs with better stability. Strikingly, bulk ultra-thin DWCNTs can be obtained with narrow diameter distribution using SWCNTs without an additional carbon source. Furthermore, we demonstrate that both metallic and semiconducting inner tubes are identified inside different types of SWCNTs. Compared to the DWCNTs obtained from the fullerene/ferrocene peapods, surprisingly the DWCNTs in our experiment has higher/similar filling ratio of inner tubes.

## Experimental and method

SWCNTs with different average diameters (see Table 1) were used in this study: high-pressure carbon monoxide (HiPco) SWCNTs with a mean diameter of 1.1 nm were purchased from Carbon Nanotechnologies Incorporated (P0169); eDIPS-SWCNTs with average diam-eter around 1.3 and 1.7 nm were prepared by enhanced direct injection pyrolytic synthesis (eDIPS) method, labeled as eDIPS-1.3 and eDIPS-1.7, respectively;[37] semiconducting and metallic SWCNTs were synthesised by arc-discharge method (marked as arc-1.4) and separated by the density gradient ultracentrifugation method.[38] Semiconducting and metallic SWCNTs were used as obtained, whereas HiPco and eDIPS SWCNTs were purified by a three-step process before the high-temperature annealing: First, hydrochloric acid (HCl) treatment was applied to remove part of the catalysts. Second, the filtered sample was annealed in air at 400 °C to remove the amorphous carbon that encapsulated catalyst particles. Third, HCl treatment again to remove the remaining catalysts and then washed out by deionised water. Buckypapers containing purified SWCNTs were obtained after filtering and drying.

In order to create the inner tubes, all of the SWCNT samples in the form of buckypaper were annealed at high temperature between 1300 and 1550 °C for 1 hour under high vacuum (below $10^{-7}$ bar). Note that the purification process before high-temperature annealing is necessary, otherwise the catalyst residues in the sample would destroy the SWCNTs significantly.

For comparison, CVD DWCNTs with average outer diameter of 1.5 nm were synthesized by high-vacuum CVD.[11] Another type of DWCNTs was prepared from ferrocene filled eDIPS-1.3 SWCNTs.[39,40] In short, purified eDIPS-1.3 SWCNTs opened by an air treatment at 400 °C were sealed together with ferrocene in a quartz tube under vacuum below $10^{-6}$ mbar. In order to fill ferrocene inside the SWCNTs, the sealed sample was heated at 400 °C for 2 days. Then the ferrocene inside eDIPS-1.3 SWCNTs was transformed into inner tubes at 800 °C in a furnace under vacuum below $10^{-6}$ mbar for 1 hour. Similarly, $C_{60}$) was filled inside arc-1.4 SWCNTs and transformed into DWCNTs in vacuum at 1250 °C.

The samples were measured by Raman spectroscopy (LabRAM, Horiba) with 458, 588, 514, 531, 568, and 633 nm excitations, as well as Fourier transform Raman spectroscopy (MultiRAM, Bruker) with a 1064 nm laser. For easy comparison, all the Raman spectra were normalized to the intensity of G-band. HRTEM (JEOL 2010F) was performed at 120 KV to observe the formation of the inner tubes inside SWCNTs. The semiconducting and metallic SWCNTs were dispersed in toluene by tip-ultrasonication for 30 min. The dispersed solution samples and the toluene (as background) were measured by optical absorption in the region of 400-1200 nm (Bruker VERTEX 80v FTIR spectrometer with spectral resolution of better than 0.2 $cm^{-1}$).

## Results and discussion

Table 1: Information on the samples before and after annealing. The area of inner tubes for the annealed Hipco sample is obtained from the spectrum excited by a 633 nm laser.

| Sample | Diameter (nm) | Area of inner tubes (1064/568 nm) | D/G ratio (pristine/annealed |
|---|---|---|---|
| eDIPS-1.7 SWCNTs | 1.7 | 4.1 / 1.9 | 0.022 / 0.029 |
| eDIPS-1.3 SWCNTs | 1.3 | 4.5 / 5.0 | 0.021 / 0.029 |
| Ferrocene@eDIPS-1.3 | 1.3 | 4.7 / 5.1 | 0.021 / 0.023 |
| $C_{60}$@arc-1.4 | 1.3 | 3.0 / 2.9 | 0.015 / 0.018 |
| CVD DWCNTs | 1.6 (outer tube) | 5.0 / 6.9 | 0.026 / 0.027 |
| Metallic SWCNTs | 1.5 | 2.7 / 3.3 | 0.030 / 0.022 |
| Semiconducting SWCNTs | 1.4 | 4.9 / 1.9 | 0.066 / 0.104 |
| HiPco SWCNTs | 1.1 | 0.3 (633 nm) | 0.008 / 0.018 |

Different high-purity SWCNTs have been used in this study, as summarized in Table 1. The D/G ratios of most of pristine samples are smaller than 0.03, except the one for semiconducting tubes, whose D/G ratio is 0.066, suggesting that they were partially damaged by the separation process[38,41]. After high-temperature annealing at 1500 °C the D/G ratios of the samples slightly increase, but in all cases remaining at low values, reflecting that the graphitisation of the annealed samples keeps very high and defect density is still low.

Deepening into the analysis of the Raman spectra, as expected for the eDIPS-1.7 sample, the peak located at around 120 $cm^{-1}$ corresponding to the SWCNTs with diameter of about 2.1 nm in the radial breathing mode (RBM) region down-shifts to lower frequency with the increased annealing temperature (see Figure 1a), indicating that these large SWCNTs are indeed enlarged, which is consistent with previous studies[28–30]. When the annealing temperature is above 1480 °C, the D'-band appears due to reconstruction of the nanotubes during the enlargement (Fig. 1b) and the 2D-band shifts to higher frequency with the increased temperature (the larger the nanotube, the higher the frequency of 2D mode[42]). It is more interesting that in the RBM region the peaks located between 150 and 170 cm$^{-1}$ shift to higher frequencies corresponding to smaller tubes and many new sharp peaks, which belong to ultra-thin nanotubes appear even starting at 1300 °C (earlier than the nanotube enlargement occurs) and maximising at 1500 °C, as seen in Fig. 1a (excited by a 1064 nm laser). This can also be seen in Figures S1 and S2 in the supplementary information showing multiple frequencies study. Above 1550 °C the DWCNTs are damaged and transformed into larger tubes or MWCNTs, as always reported previously. Hence, the temperature window for the synthesis of inner tubes is quit narrow, mainly between 1400 and 1550 °C, which prevents the formation of inner tubes to be observed previously. With the appearance of these newly-formed and up-shifted RBM peaks, the 2D-band shifts to lower frequency (Fig. 1c)[42]. The diameter of the nanotubes can be calculated by the formula D = 234/(RBM-10)[43], where D is the diameter of nanotube in nm and RBM corresponds to the Raman shift of RBM peaks in cm$^{-1}$. Therefore, the calculated diameters of these new RBM peaks between 260 and 360 cm$^{-1}$ belong to tubes from 0.65 to 0.96 nm. Note that no small SWCNTs exist in the pristine sample inspecting it with different laser lines, as shown in the Fig. 1a and Fig. S2b in the supplementary information. Hence, this is the first hint toward the formation of inner ultra-thin tubes inside the pristine SWCNTs. Another evidence of inner tube formation is that the diameter difference between the tubes with up-shifted RBM peaks and the new-formed small tubes is in the range of 0.65-0.75 nm, which is about twice the interplanar distance of graphite, implying that they are paired as outer and inner tubes, respectively. The carbon sources for the formation of the inner tubes are: First, the residue carbon and amorphous carbon inside and/ or surrounded the nanotubes; Second, the additional carbon atoms from the SWCNTs when they are reorganized into smaller SWCNTs during the annealing, evidenced by the up–shifted RBM peaks as shown in Fig. 1a. The new-formed inner tubes sometimes are defective and a few of them are even not integral, as seen the TEM image of Figure 3(d), which explains the increased D/G ratio. In addition, the reformation of the SWCNTs during the annealing, which creases defects as well, as shown in Fig. 1. At higher temperature, those defects could be partly healed, which is similar as the reported phenomenon [25-27]. However, in general the D/G ratios of all the samples are quite low.

We also did the annealing experiments on eDIPS-1.3 and HiPco SWCNTs. Similar results have been obtained for the eDIPS-1.3 samples that inner tubes can be synthesized inside pristine SWCNTs. A comparison between the pristine sample and the same annealed at 1500 °C is shown in Fig. 2. Further information can also be found in the supplementary information Figures S3 and S4. For the HiPco SWCNTs, the diameter enlargement was clearly observed in the Raman spectra excited by 568 and 633 nm lasers (Figure S5). A general observation for Hipco tubes is that only a small amount of SWCNTs has been converted into thin DWCNTs, because most of the HiPco SWCNTs are too thin to be used as reactors for the synthesis of unstable tubes below 0.5 nm with great curvature and huge strain. A possible case can be seen in Fig. 2, showing additional RBM peaks belong to (7,2) and (8,0) tubes at around 360 cm$^{-1}$ excited by 633 nm laser appear with annealing, which may be pared with the down-shifted RBM peaks (enlarged tubes) at around 190 cm$^{-1}$.

In order to confirm the Raman spectroscopic results on the inner tube formation, high-resolution transmission electron microscopy (HRTEM) was carried out. Fig. 3 and Figs. S7 and S8 show several high magnification images of the eDIPs 1.3nm tubes annealed at 1500 °C, where the inner tube formation is evident. The new created DWCNTs are mostly with well–defined structure of inner tubes, as shown in Figs. 3a-3c. However, we still can find a few SWCNTs with segments of inner tubes (Fig. 3d). Statistically, the ratio of the DWCNTs in the sample is higher than 90 % (Fig. 3e), which is similar as the one reported on the CVD-DWCNTs [7,9,11]. As seen in Fig. 3f, the diameter distributions of the inner and outer tubes from many TEM images for the eDIPS-1.3 sample indicate that the sample contains a large population of ultra-thin DWCNTs, which is in line with our Raman spectroscopic results. Therefore, compared to the CVD and arc-discharge methods it is a big advantage for our high-temperature annealing method to obtain ultra-thin DWCNTs, where the outer diameter of such ultra-thin DWCNTs mostly depends on the pristine SWCNTs. As an example, super small inner tubes, *e.g.* (5,3) and (6,1) tubes located from 350 to 440 cm$^{-1}$ were surprisingly identified by the Raman spectra of samples annealed above 1460 °C, as seen in Figures S1 and S2. Although TEM observations have revealed the presence of very large SWCNTs in our samples, no large DWCNTs could be identified, which is in keeping with the Raman results. In this case, no additional carbon source has been used, which suggests that the absence of large inner tubes requires a carbon supply.

For comparison, DWCNTs were also obtained by transforming ferrocene ( or *C*$_{60}$) filled inside eDIPS-1.3 SWCNTs as additional carbon source into the tubes at 900 (or 1250) °C[39,40]. In such cases, additional RBM peaks from 250 to 360 cm$^{-1}$ corresponding to inner tubes were observed after the transformation (Fig. 4). Raman spectra of transformed transformed *C*$_{60}$@arc-discharge SWCNTs, ferrocene@eDIPS-1.3, annealed eDIPS-1.3, annealed eDIPS-1.7, and CVD DWCNTs excited by 1064 and 568 nm lasers are displayed with offsets in Fig. 4. Interestingly, no significant difference among these samples could be seen in the Raman spectra excited at 1064 nm, although the inner tubes of the annealed SWCNTs are in principle not as perfect as the CVD DWCNTs (without D'-band) due to the limited carbon source available for the inner tube formation. Introducing ferrocene allows inner tube formation with less defects (no D'-band) at much lower temperature, but the number of RBM peaks and their intensity is not increased too much compared to that of the annealed eDIPS-1.3 sample. Moreover, since the CVD sample consists of above 90 % of DWCNTs[11], the transformed or the annealed samples also contain similar ratio of DWCNTs, when a nanotube is considered as DWCNT if there is an inner tube inside host SWCNTS independent from its length. This was statistically confirmed by the TEM image analysis. For the annealed eDIPS-1.3 and 1.7 samples, the carbon sources are partly from the amorphous carbon, which always exists in the sample, and partly from the remaining carbon due to the nanotube reconstruction during high-temperature annealing: Either from the enlarged carbon nanotubes, *i.e.*, coalescence of two nanotubes (down-shifted RBM peaks corresponding to larger tube formation), or from the size-reduced carbon nanotube (the up-shifted RBM peaks corresponding to the outer tubes).

Since there are two groups of SWCNTs, *i.e.* metallic and semiconducting, it is important to clarify whether there is difference or not when synthesising the inner tubes inside these two types of SWCNTs. Separated metallic and semiconducting SWCNTs with corresponding high purities greater than 95 % and 99 % were annealed at 1500 °C. Their purity could be confirmed from their absorption spectra as seen in the Fig. S7. The results show that although the metallic tubes had much larger D-band (Fig. 5), apparently such difference does not have much influence on the inner tube formation because RBM peaks look similar between these two annealed samples. Since the average diameter of the metallic tubes is slightly larger than that of the semiconducting tubes, the new formed inner tubes inside metallic tubes are also a little bit larger, as seen the RBM peaks excited by different lasers in Figs. S8 and S9. Compared to the other annealed samples, more thin metallic inner tubes can be formed inside these separated SWCNTs if carefully looking at the Raman spectra of all the annealed samples excited by 458 nm laser, simply because separated tubes with diameter 1.4-1.5 nm are more suitable for the synthesis of such metallic tubes with diameter of 0.7-0.9 nm.

For detailed analysis of the inner tube formation, we compare the area of inner tubes in the Raman spectra excited by two lasers to statistically analyze the difference among the samples. The results are summarized in Table 1. The CVD DWCNTs have the highest inner tubes area in both cases as expected. Surprisingly, the area of annealed eDIPS-1.3 sample is comparable to the area of transformed ferrocene@eDIPS-1.3 sample, and both of them are much higher than that of $C_{60}$)@arc-1.4 sample, whereas no identified inner tube formation is found for the Hipco sample if excited by 568 and 1064 nm lasers. In addition, both of the annealed semiconducting and metallic samples represent clearly inner tube formation, confirming again that the conductivity type of the hosting SWCNTs does not affect the inner tube's synthesis. In Table 1, it is suggested that D/G ratio is not related to inner tube formation, but the diameter of the host SWCNTs is the most important factor for the inner tube synthesis. The reasons are obvious: If the SWCNTs are too small like the HiPco SWCNTs, formation of unstable ultra-thin inner tubes is more difficult than the formation of larger ones; If the SWCNTs are too large, formation of big inner tubes would need more residual carbon atoms for the construction of inner tubes, which usually is not enough in the sample unless supplying additional carbon source from the filled molecules, *e.g.* $C_{60}$[22,23] or ferrocene[39,40].

Finally, we discuss on the conductivity type of the new formed inner tubes. In order to directly see the difference between different samples, the diameters of the new formed inner tubes were extracted from RBM peaks measured with the 458, 488, 514, 568, 633, and 1064 nm laser excitation wavelengths allowing to access the first optical transition energy ($E_{11}$) of metallic tubes as well as the first and second optical transition energies ($E_{11}$ and $E_{22}$) of semiconducting ones. Following the Kataura plots[44], the semiconducting thin nanotubes would be excited by most of the visible lasers used in this study. Therefore, semiconducting inner nanotubes with diameter of 0.55-0.95 nm can always be observed in our Raman spectra, as seen in Figures 1 and 2-5 as well as Figures S1-S5. Metallic inner tubes with larger diameter from 0.85 to 1.20 nm on the other hand would be in resonance with 458-532 nm lasers in this study, *e.g.*, in Figure S2 a weak RBM peak at 280 cm$^{-1}$ excited by 458 nm laser is assigned to (6,6) tube, the RBM peak at around 250 cm$^{-1}$ excited by 514 nm laser are belong to Family21 (2n+m=21) tubes, and the RBM peaks at 210-250 cm$^{-1}$ excited by 568 nm laser are from Family24 tubes. More details can be seen in the statistics as shown in Figure 6. The results are summarized into two groups, *i.e.* metallic and semiconducting inner tubes. In general, both semiconducting and metallic inner tubes can be synthesized in all of the SWCNTs and ultra-thin DWCNTs can be obtained by high-temperature annealing.

From current measurement conditions, for all of the samples the diameter of metallic tubes is always slightly larger than the semiconducting ones. Smaller metallic inner tubes could also exist in the sample, however, in this case ultraviolet laser should be used for detection. Also, there is a clear trend that the median diameter of new created inner tubes depends on the diameter of the host SWCNTs, because the diameter different between the outer and the inner tubes should roughly match twice the interplanar distance of graphite. In addition, the DWCNTs with large inner tubes are not considered when analyzing the Raman spectra, because the overlapping of the signal from the large inner tubes and from small outer tubes makes the analysis very difficult. Yet, as shown in Fig. 3f, statistics of the inner and outer tubes from many TEM images for the eDIPS-1.3 sample fits quit well with the results obtained by multi-frequency Raman spectroscopy.

## Conclusions

We have shown that high temperature high vacuum annealing as a universal route to synthesize high-yield DWCNTs with narrow diameter distribution using SWCNTs as starting materials. The inner tube formation was found only within a very narrow window, otherwise the inner tubes are damaged again at higher temperature. The new formed inner tubes are ultra-thin, for example even (6,1), (5,3), (7,2), and (8,0) exist in the annealed samples. Similar results have been found when applying metallic and semiconducting SWCNTs as the host nanotubes for inner tube formation. Also, both semiconducting and metallic inner tubes can be synthesized in all SWCNTs. Applying this new post-treatment route, ultra-thin DWCNTs can be easily and fast prepared in bulk without additional carbon source. Also, the filling ratio of inner tubes is higher than (comparable to) the one obtained from the transformed fullerene (ferrocene)@SWCNTs. Combining the advantages of both SWCNTs and MWCNTs, such high-quality DWCNTs can be applied to most of the applications explored using SWCNTs and/or MWCNTs.

# Experimental and method

SWCNTs with different average diameters (see Table 1) were used in this study: HiPco SWCNTs with a mean diameter of 1.1 nm were purchased from Carbon Nanotechnologies Incorporated (P0169); eDIPS-SWCNTs with average diameter around 1.3 and 1.7 nm were prepared by enhanced direct injection pyrolytic synthesis (eDIPS) method, labeled as eDIPS-1.3 and eDIPS-1.7, respectively[37]; semiconducting and metallic SWCNTs were synthesised by arc-discharge method (marked as arc-1.4) and separated by the density gradient ultra-centrifugation method [38]. Semiconducting and metallic SWCNTs were used as obtained, whereas HiPco and eDIPS SWCNTs were purified by a three-step process before the high-temperature annealing: First, hydrochloric acid (HCl) treatment was applied to remove part of the catalysts. Second, the filtered sample was annealed in air at 400 °C to remove the amorphous carbon that encapsulated catalyst particles. Third, HCl treatment again to remove the remaining catalysts and then washed out by deionised water. Buckypapers containing purified SWCNTs were obtained after filtering and drying.

In order to create the inner tubes, all of the SWCNT samples in the form of buckypaper were annealed at high temperature between 1300 and 1550 °C for 1 hour under high vacuum (below $10^{-7}$ bar). Note that the purification process before high-temperature annealing is necessary, otherwise the catalyst residues in the sample would destroy the SWCNTs significantly.

For comparison, CVD DWCNTs with average outer diameter of 1.5 nm were synthesized by high-vacuum CVD[11]. Another type of DWCNTs was prepared from ferrocene filled eDIPS-1.3 SWCNTs [39,40]. In short, purified eDIPS-1.3 SWCNTs opened by an air treatment at 400 °C were sealed together with ferrocene in a quartz tube under vacuum below $10^{-6}$ mbar. In order to fill ferrocene inside the SWCNTs, the sealed sample was heated at 400 °C for 2 days. Then the ferrocene inside eDIPS-1.3 SWCNTs was transformed into inner tubes at 800 °C in a furnace under vacuum below $10^{-6}$ mbar for 1 hour. Similarly, $C_{60}$) was filled inside arc-1.4 SWCNTs and transformed into DWCNTs in vacuum at 1250 °C.

The samples were measured by Raman spectroscopy (LabRAM, Horiba) with 458, 588, 514, 531, 568, and 633 nm excitations, as well as Fourier transform Raman spectroscopy (MultiRAM, Bruker) with a 1064 nm laser. For easy comparison, all the Raman spectra were normalized to the intensity of G-band. HRTEM (JEOL 2010F) was performed at 120 KV to observe the formation of the inner tubes inside SWCNTs. The semiconducting and metallic SWCNTs were dispersed in toluene by tip-ultrasonication for 30 min. The dispersed solution samples and the toluene (as background) were measured by optical absorption in the region of 400-1200 nm (Bruker VERTEX 80v FTIR spectrometer with spectral resolution of better than 0.2 $cm^{-1}$).

# Acknowledgement

This work was supported by the Austrian Science Funds (FWF, P27769-N20) and the European Union's Horizon 2020 research and innovation programme under grant agreement No 664878.

# Supporting Information Available

Raman spectra of pristine/annealed eDIPS-1.7/1.3 nm, HiPco SWCNTs, and semiconducting/metallic SWCNTs. TEM image of annealed eDIPS-1.3 nm SWCNTs. Absorption of semiconducting/metallic SWCNTs.

**Graphical TOC Entry**

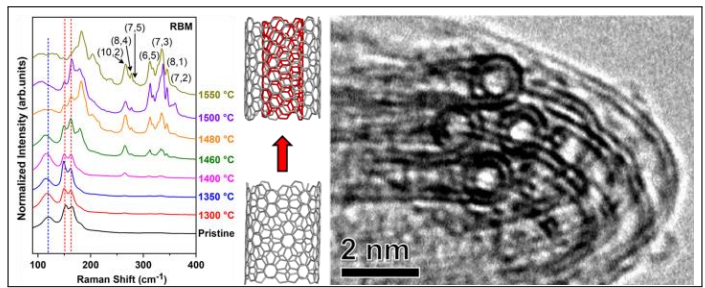

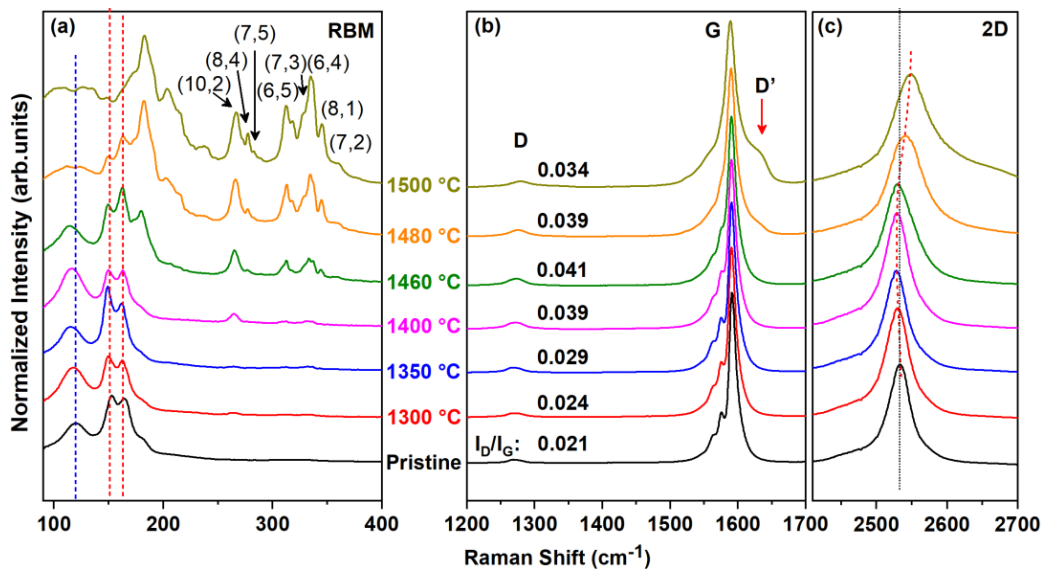

Figure 1: Raman spectra of pristine and annealed eDIPS-1.7 nm SWCNTs at different temperatures excited by a 1064 nm laser. (a) RBM, (b) D and G–bands, (c) 2D–band.

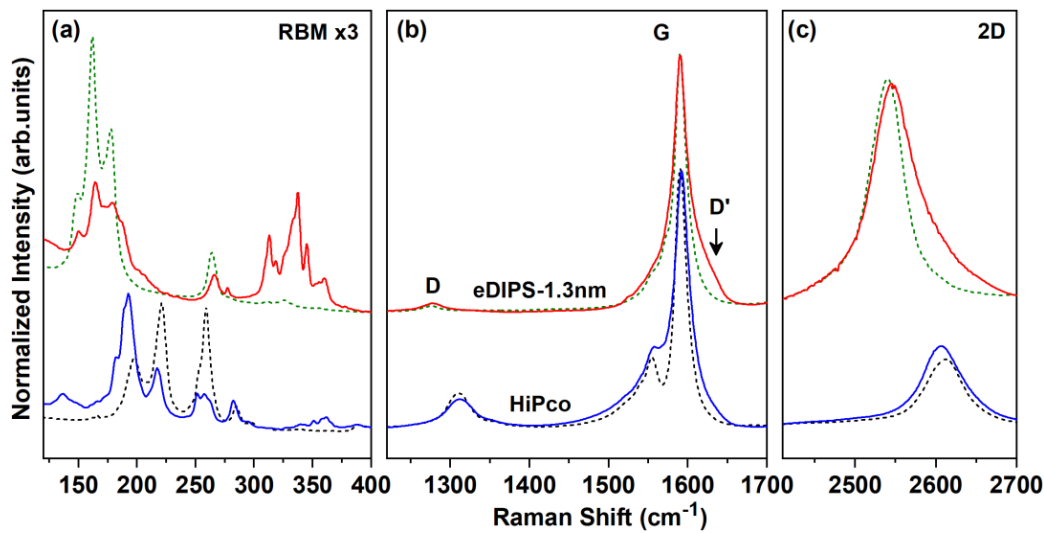

Figure 2: Raman spectra of pristine (dashed lines) and annealed (solid lines) eDIPS-1.3 nm (excited at 1064 nm) and HiPco SWCNTs (excited at 633 nm). (a) RBM, (b) D and G–bands, (c) 2D–band.

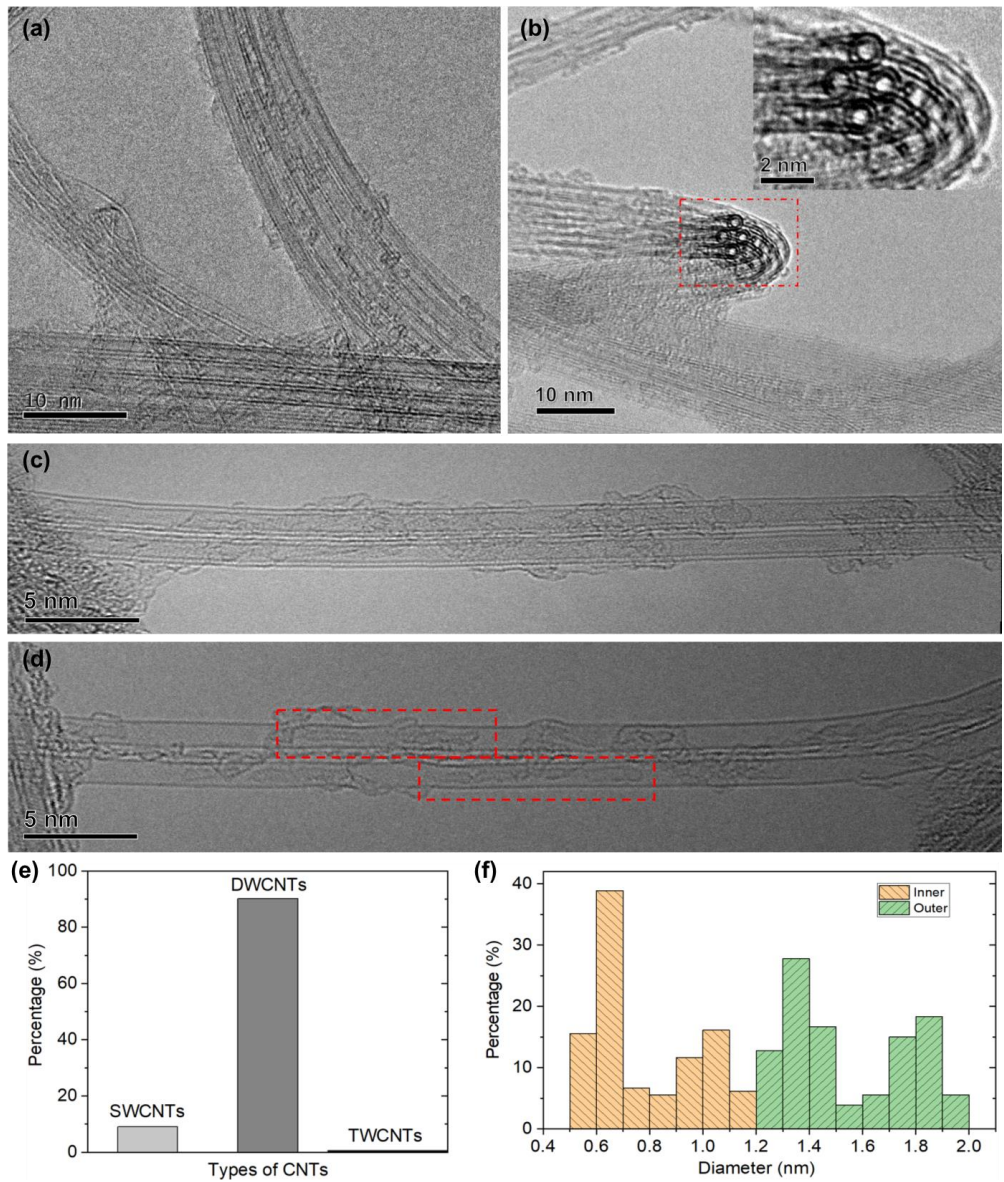

Figure 3: HRTEM images of annealed SWCNTs at 1500 °C. (a) Bundles of DWCNTs in annealed eDIPS-1.3. (b) The cross section of a bundle of DWCNTs in annealed eDIPS-1.3. The observed DWCNTs have similar diameters within a narrow distribution. Inset of (b): Amplified cross section of a bundle of DWCNTs. Two SWCNTs with integral (c) and non-integral (d) inner tubes. Statistics of types of CNTs (e) and diameter distribution of DWCNTs (f) in the sample.

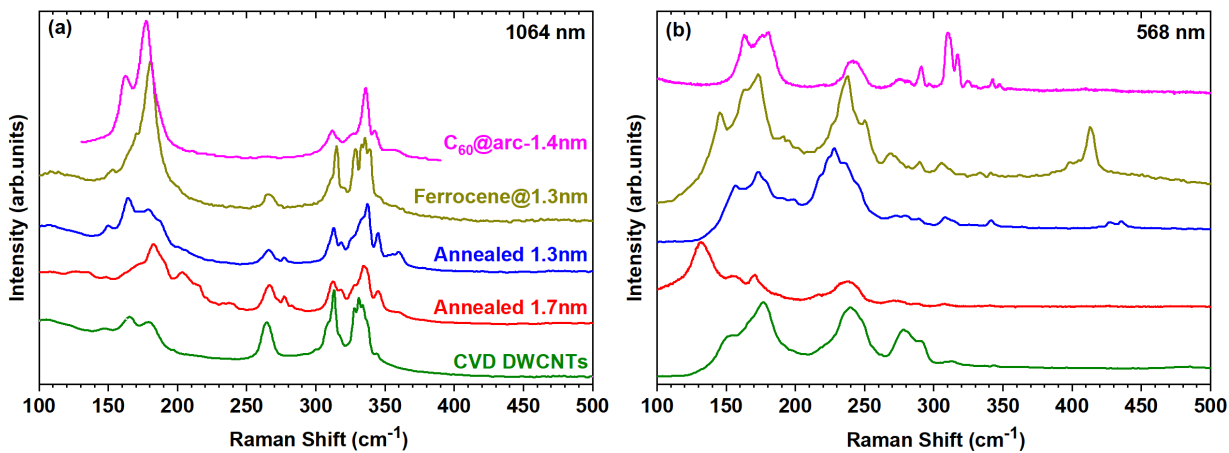

Figure 4: Comparison of Raman spectra of CVD-grown DWCNTs, annealed eDIPS-1.7 nm SWCNTs, annealed eDIPS-1.3 nm SWCNTs, transformed ferrocene@eDIPS-1.3 nm SWCNTs, and transformed ferrocene@arc-discharge-1.4 nm SWCNTs excited by (a) 1064 nm and (b) 568 nm laser wavelengths.

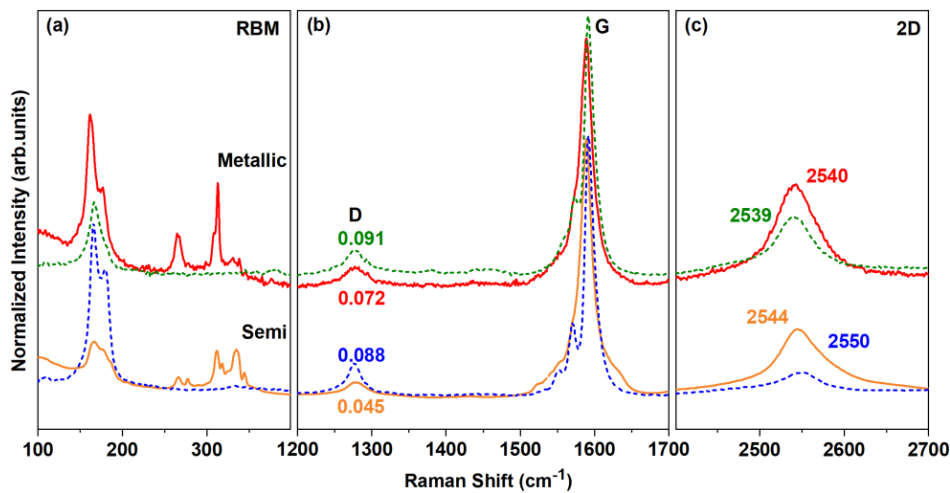

Figure 5: Raman spectra of pristine/annealed (dashed/solid lines) metallic and semiconducting arc-discharge SWCNTs excited by a 1064 nm laser wavelength. (a) RBM, (b) D and G–bands, (c) 2D–band.

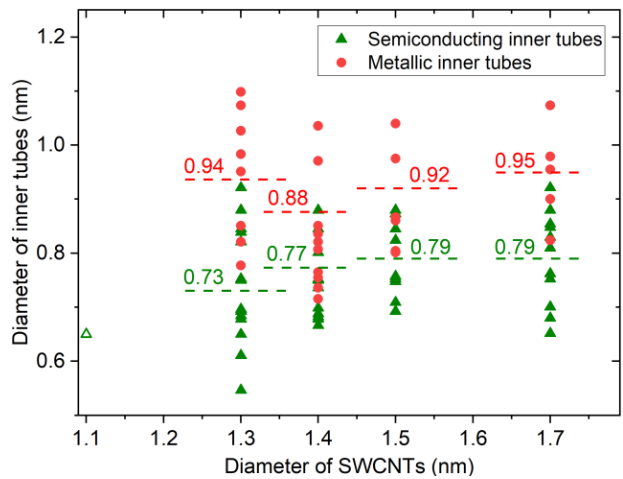

Figure 6: Statistics of the diameter distribution. Diameter of inner tubes extracted from the RBM peaks measured by multi-frequency Raman spectroscopy. The dashed horizontal lines indicate the median diameter for each sample as a guide to the eye.